\newif\ifBibtex

\documentclass[ALICE,manyauthors]{cernphprep}
\usepackage{amssymb}
\usepackage{graphicx}
\usepackage{epstopdf}
\usepackage{lineno}
\usepackage{hyperref}
\newcommand{\CommentBlock}[1]{}

\newcommand{\PbPb}{Pb--Pb}
\newcommand{\pPb}{p--Pb}

\newcommand{\pp}{pp}

\newcommand{\sqrtsNN}{\ensuremath{\sqrt{s_{\mathrm {NN}}}}}
\newcommand{\sqrts}{\ensuremath{\sqrt{s}}}

\newcommand{\rr}{\ensuremath{R}}

\newcommand{\invnb}{{nb}$^{-1}$}
\newcommand{\gev}{GeV/$c$}

\newcommand{\antikT}{anti-$k_{ \mathrm T}$}

\newcommand{\pT}{\ensuremath{p_\mathrm{T}}}
\newcommand{\pTjet}{\ensuremath{p_{\mathrm{T,jet}}}}

\newcommand{\kzerol}{\ensuremath{\mathrm{K}^{0}_\mathrm{L}}}
\newcommand{\kzeros}{\ensuremath{\mathrm{K}^{0}_\mathrm{S}}}

\newcommand{\zleading}{\ensuremath{z_\mathrm{leading}}}
\newcommand{\pTlow}{\ensuremath{p_{\mathrm T}^{\mathrm{low}}}}
\newcommand{\pThigh}{\ensuremath{p_{\mathrm T}^{\mathrm{high}}}}
\newcommand{\CMC}{\ensuremath{C_{\mathrm{MC}}}}
\newcommand{\CMCwithArgs}{\ensuremath{C_{\mathrm{MC}}\left(\pTlow;\pThigh\right)}}

\newcommand{\dfmeasdpT}{\ensuremath{\frac{\mathrm{d}F^{\mathrm {uncorr}}_{\mathrm {meas}}}{\mathrm{d} \pT}}}

\newcommand{\dsigMCpart}{\ensuremath{\mathrm{d} \sigma_{\mathrm{MC}}^{\mathrm {particle}}/ \mathrm{d} \pT}}
\newcommand{\dsigMCdet}{\ensuremath{\mathrm{d} \sigma_{\mathrm{MC}}^{\mathrm {detector}}/\mathrm{d} \pT}}

\newcommand{\Eclust}{\ensuremath{E_{\rm{clust}}}}
\newcommand{\Ecorr}{\ensuremath{E_{\rm{corr}}}}
\newcommand{\sump}{\ensuremath{\Sigma_{\rm{p}}}}
\newcommand{\fsub}{\ensuremath{f_{\rm{sub}}}}

\newcommand{\pTpart}{\ensuremath{p_\mathrm{T,jet}^\mathrm{particle}}}
\newcommand{\pTdet}{\ensuremath{p_\mathrm{T,jet}^\mathrm{detector}}}

\begin{document}

\begin{titlepage}
\PHnumber{2012-301}                 
\PHdate{Oct 10, 2012}              




\title{Measurement of the inclusive differential jet cross section\\ in \pp\ collisions at \sqrts\ = 2.76 TeV}
\ShortTitle{Inclusive jet cross section in pp collisions at \sqrts\ = 2.76 TeV}   
\Collaboration{ALICE Collaboration~\thanks{See Appendix~\ref*{app:collab} for the list of collaboration members}}
\ShortAuthor{ALICE Collaboration}      



\begin{abstract} 
The ALICE collaboration at the CERN Large Hadron Collider reports the first measurement of the inclusive differential jet cross section at mid-rapidity in \pp\ collisions at \sqrts\ = 2.76 TeV, with integrated luminosity of 13.6 \invnb. Jets are measured over the transverse momentum range 20 to 125 \gev\ and are corrected to the particle level. Calculations based on Next-to-Leading Order perturbative QCD are in good agreement with the measurements. The ratio of inclusive jet cross sections for jet radii \rr\ = 0.2 and \rr\ = 0.4 is reported, and is also well reproduced by a Next-to-Leading Order perturbative QCD calculation when hadronization effects are included.
\end{abstract}
\end{titlepage}

\maketitle
\setcounter{page}{2}

\section{Introduction}

A QCD jet is a collimated shower of particles arising from the hadronization of a highly virtual quark or gluon generated in a hard (high momentum transfer $Q^2$) scattering. Perturbative Quantum Chromodynamics (pQCD) calculations of inclusive jet cross sections agree with collider measurements over a wide kinematic range, for a variety of collision systems \cite{Aad:2011fc,CMS:2011ab,Abulencia:2007ez,Aaltonen:2008eq,Abazov:2008ae}. Jets provide important tools for studying Standard Model and Beyond Standard Model physics, as well as hot and dense QCD matter that is created in high energy collisions of heavy nuclei. In heavy-ion collisions, large transverse momentum (\pT) partons traverse the colored medium and lose energy via induced gluon radiation and elastic scattering, which modify jet structure relative to jets generated in vacuum. These modifications (``jet quenching'') may be observable experimentally, and can be calculated theoretically (\cite{Majumder:2010qh} and references therein).

Measurements of the properties of the hot and dense QCD medium generated in \PbPb\ collisions at the Large Hadron Collider (LHC) require reference data from more elementary collisions (\pp\ and \pPb), in which generation of a QCD medium is not expected. In March 2011, the LHC undertook a three-day run with \pp\ collisions at \sqrts\ = 2.76 TeV, the same center-of-mass energy as the currently available \PbPb\ data, to obtain first measurements of such reference data. This paper reports the measurement of the inclusive differential jet cross section at mid-rapidity from that run, based on integrated luminosity of 13.6 \invnb.

Jet reconstruction for this analysis utilizes the infrared-safe and collinear-safe \antikT\ algorithm \cite{Cacciari:2008gp,Cacciari:2011ma}. The algorithm requires specification of a clustering parameter \rr, which is the maximum distance in pseudorapidity $\eta$ and azimuthal angle $\varphi$ over which constituent particles are clustered, $\sqrt{(\Delta\eta)^2+(\Delta\varphi)^2}<\rr$. We study the dependence of the inclusive jet cross section on \rr, which is sensitive to the transverse structure of jets, and compare our measurements to pQCD calculations at Next-to-Leading Order (NLO) \cite{Frixione:1995ms,Frixione:1997np,Armesto,Soyez:2011np}.


\section{Detector and Data Set}

ALICE consists of two large-acceptance spectrometers \cite{Aamodt:2008zz}: the central detector, containig a high precision tracking system, particle identification detectors, and calorimetry, all located inside a large solenoidal magnet with field strength 0.5 T; and a forward muon spectrometer. Only the central detector is used for this analysis.

The data were recorded by the ALICE detector for \pp\ collisions at \sqrts\ = 2.76 TeV. Several trigger detectors were utilized: the VZERO, consisting of segmented scintillator detectors covering the full azimuth over $2.8<\eta<5.1$ (VZERO-A) and $-3.7<\eta<-1.7$ (VZERO-C); the SPD \cite{Aamodt:2010aa}, a two-layer silicon pixel detector consisting of cylinders at radii 3.9 cm and 7.6 cm from the beam axis and covering the full azimuth over $|\eta|<2$ and $|\eta|<1.4$ respectively; and the EMCal \cite{Cortese:2008zza,Allen:2009aa}, an Electromagnetic Calorimeter covering 100 degrees in azimuth and $|\eta|<0.7$. The EMCal for this measurement consists of 10 supermodules with a total of 11520 individual towers, each covering an angular region $\Delta\eta\times\Delta\varphi=0.014\times0.014$. The EMCal Single Shower (SSh) trigger system generates a fast energy sum (800 ns) at Trigger Level 0 (L0) for overlapping groups of $4\times4$ ($\eta\times\varphi$) adjacent EMCal towers, followed by comparison to a threshold energy. Event recording was initiated by two different trigger conditions: (i) the Minimum Bias (MB) trigger, requiring at least one hit in any of VZERO-A, VZERO-C, and SPD, in coincidence with the presence of an LHC bunch crossing, and (ii) the EMCal SSh trigger, requiring that the MB trigger condition is satisfied and that at least one SSh sum exceeds a nominal threshold energy of 3.0 GeV. The MB trigger cross section was measured to be $55.4\pm1.0$ mb by a van der Meer scan \cite{2012sja}.

The primary event vertex was reconstructed as described in \cite{Abelev:2012hxa}. Events selected for offline analysis were required to have a reconstructed primary vertex within 10 cm of the center of the ALICE detector along the beam axis. After event selection cuts, the MB-triggered data set corresponds to integrated luminosity of 0.5 \invnb, while the EMCal-triggered data set corresponds to 13.1 \invnb.

Simulations are based on the PYTHIA6 \cite{Sjostrand:2006za} (Perugia-2010 tune, version 6.425) and HERWIG \cite{HERWIG} (version 6.510) Monte Carlo event generators. ``Particle-level'' simulations utilize the event generator output directly, without accounting for detector effects, while ``detector-level'' simulations also include a detailed particle transport and detector response simulation based on GEANT3 \cite{GEANT3}. 

For offline analysis, input to the jet reconstruction algorithm consists of charged particle tracks and EMCal clusters. Charged particle tracks are measured in the ALICE tracking system, which covers the full azimuth within $|\eta|<0.9$. The tracking system consists of the ITS \cite{Aamodt:2010aa}, a high precision, highly granular Inner Tracking System consisting of six silicon layers including the SPD, with inner radius 3.9 cm and outer radius 43.0 cm, and the TPC \cite{Alme:2010ke}, a large Time Projection Chamber with inner radius 85 cm and outer radius 247 cm, that measures up to 159 independent space points per track. 

In order to achieve high and azimuthally uniform tracking efficiency required for jet reconstruction, charged track selection utilizes a hybrid approach that compensates local inefficiencies in the ITS. Two distinct track classes are accepted in the hybrid approach: (i) tracks containing at least three hits in the ITS, including at least one hit in the SPD, with momentum determined without the primary vertex constraint, and (ii) tracks containing less than three hits in the ITS or no hit in the SPD, with the primary vertex included in the momentum determination. Class (i) contains 90\%, and class (ii) 10\%, of all accepted tracks, independent of \pT. Track candidates have Distance of Closest Approach to the primary vertex less than 2.4 cm in the plane transverse to the beam, and less than 3.0 cm in the beam direction. Accepted tracks have measured $\pT>0.15$ \gev, with a \pT-dependent minimum number of space points in the TPC ranging from 70 at $\pT=0.15$ \gev\ to 100 for $\pT>20$ \gev. Tracking efficiency for charged pions from the primary vertex is approximately $60 \%$ at \pT\ = 0.15 \gev, increasing to about $87\%$ for $3<\pT<40$ \gev. Charged track momentum resolution is estimated on a track-by-track basis using the covariance matrix of the track fit \cite{Fruhwirth:1987fm}, and is verified by the invariant mass resolution of reconstructed $\Lambda$ and \kzeros\ \cite{Abelev:2012hxa}. The momentum resolution $\delta\pT/\pT$ is approximately $1\%$ at \pT\ = 1.0 \gev\ and approximately $4\%$ at \pT\ = 40 \gev\ for track class (i) and approximately $1\%$ at \pT\ = 1.0 \gev\ and approximately $7\%$ at \pT\ = 40 \gev\ for track class (ii). Charged tracks with $\pT>40$ \gev\ make negligible contribution to the inclusive jet population considered in this analysis.

EMCal clusters are formed by a clustering algorithm that combines signals from adjacent EMCal towers, with cluster size limited by the requirement that each cluster contains only one local energy maximum. A noise suppression threshold of 0.05 GeV is imposed on individual tower energies, and the cluster energy must exceed 0.3 GeV. Noisy towers, identified by their event-averaged characteristics and comprising about 1\% of all EMCal towers, are removed from the analysis. Clusters with large apparent energy but anomalously small number of contributing towers are attributed to the interaction of slow neutrons or highly ionizing particles in the avalanche photodiode of the corresponding tower, and are removed from the analysis. EMCal non-linearity was measured with test beam data to be negligible for cluster energy between 3 GeV and 50 GeV, with more energetic clusters making negligible contribution to the inclusive jet population considered in this analysis. A non-linearity correction is applied for clusters with energy below 3 GeV, with value approximately 7\% at 0.5 GeV.

Charged hadrons deposit energy in the EMCal, most commonly via minimum ionization, but also via nuclear interactions generating hadronic showers, while electrons deposit their full energy in the EMCal via electromagnetic showering. Both charged hadron and electron contributions to EMCal cluster energy are accounted for, in order not to double-count a fraction of their energy in the measured jet energy. The correction procedure, which is similar in nature to ``Particle Flow'' algorithms for jet reconstruction \cite{CMS:2010eua}, minimizes dependence of the analysis on the simulation of hadronic and EM showers. Measured charged particle trajectories are propagated to the EMCal \cite{Cortese:2008zza}, with each track then matched to the nearest cluster within $\Delta\eta = 0.015$ and $\Delta\varphi=0.03$. Multiple charged hadrons can be matched to a single cluster, though the probability for this is less than 0.2\%. Test beam measurements of single charged particle interactions in the EMCal show that the probability for the EMCal shower energy to exceed the particle momentum is negligible \cite{Cortese:2008zza}. For measured cluster energy \Eclust\ and sum of momenta of all matched tracks \sump, the corrected cluster energy \Ecorr\ is set to zero if $\Eclust<\sump$; otherwise, $\Ecorr=\Eclust-\fsub*\sump$, where $\fsub=1$ for the primary analysis. This data-driven procedure accurately removes charged particle shower energy from EMCal clusters that do not have contribution from photons or untracked charged particles (i.e. those without ``cluster pileup''), which corresponds to approximately 99\% of all clusters. Correction for residual cluster pileup effects utilizes detector-level simulations based on PYTHIA6. The simulations accurately reproduce the distribution of $(\Eclust-\Ecorr)/\sump$, which corresponds closely to an in-situ measurement of the $E/p$ distribution of the EMCal in the region $E/p<1$. 

\section{Jet Reconstruction and Trigger Bias}

Jet reconstruction is carried out utilizing the FastJet \antikT\ algorithm with boost-invariant \pT\ recombination scheme \cite{Cacciari:2008gp}, and with clustering parameters \rr\ = 0.2 and 0.4. A jet is accepted if its centroid lies within the EMCal acceptance, with distance at least $R$ to the EMCal edge. The measured cross section is corrected to acceptance $|\eta|<0.5$ and $0<\varphi<2\pi$.

The charged particle tracking algorithm may misidentify low \pT\ decay daughters from secondary vertices as primary vertex tracks, and assign them a much larger \pT\ value. In addition, background in the EMCal can generate false neutral clusters with large apparent \pT, as described above. The cuts imposed at the track or cluster level to suppress such cases directly may not be fully efficient, leading to fake jets with large apparent \pTjet. However, such false high \pT\ tracks or clusters will have little additional hadronic activity in their vicinity, if they are not a part of an energetic jet. These cases are identified by examining the distribution of $z=p_{\rm {h,proj}}/p_{\rm jet}$, the magnitude of the projection of the hadron 3-momentum on the jet axis, relative to the total jet momentum. Jets whose \pT\ is carried almost entirely by a single hadron generate a peak near $z=1$ that is found to be discontinuous with the remainder of the distribution. The fake jet population due to single mis-measured tracks or clusters is therefore removed by requiring \zleading\ $< 0.98$ independent of \pTjet, where \zleading\ refers to the $z$ value of the most energetic hadron candidate in the jet. The effect of the \zleading\ cut on the inclusive jet yield is negligible for $\pTjet>10$ \gev .

Particle-level simulations based on PYTHIA6 show negligible bias in the inclusive jet cross section due to the MB trigger, for jets in the kinematic range considered here ($\pT>20$ GeV/c).

The bias imposed on the inclusive jet cross section by the EMCal SSh trigger is determined by comparing the cross sections measured with MB and SSh triggers. However, the MB data set has limited statistical reach, and a more precise determination of the SSh trigger bias for the inclusive jet yield is carried out using a data-driven approach incorporating simulations. The first step in this approach is to measure the SSh trigger efficiency for clusters by comparing the rate of SSh-triggered clusters and clusters from MB-triggered data, whose ratio reaches a plateau for cluster energy above 5 GeV. The trigger efficiency in the plateau is assumed to be 100\% for the regions in which the trigger hardware was known to be fully functional (about 90\% of the acceptance). Detector-level simulated jet events are then generated using PYTHIA6. In order to account for local variations in trigger efficiency, each EMCal cluster in a simulated event is accepted by the trigger with probability equal to the measured cluster trigger efficiency at that energy, for the supermodule in which it is located. A simulated event is accepted by the trigger if at least one EMCal cluster in the event satisfies the trigger requirement. The cumulative trigger efficiency for jets is then determined by comparing the inclusive jet spectrum for the triggered and MB populations in the simulation. The systematic uncertainty due to trigger efficiency arises from dependence on the hadronization model, which is assessed by comparing calculations incorporating the PYTHIA6 and HERWIG generators; from the uncertainty in the online trigger threshold and in the relative scaling of SSh-triggered and MB cross sections; and from the difference in SSh trigger bias for inclusive jets determined directly from data and from the alternative, data-driven approach incorporating simulations. The resulting uncertainty decreases rapidly as jet \pT\ increases.

Particles from the underlying event (UE) should not be included in the jet measurement, but their contribution cannot be discriminated on an event-wise basis. Correction for the UE contribution was therefore applied on a statistical basis. The UE transverse momentum density was estimated to be $2.1\pm0.4$ \gev\ per unit area, using dijet measurements \cite{CDFUE} over a limited kinematic range, supplemented by PYTHIA6 particle-level simulations. The systematic uncertainty is taken as the difference in UE density between data and simulations. The corresponding uncertainty in JES for \rr\ = 0.4 jets is 1\% at \pT\ = 25 \gev\ and 0.3\% at \pT\ = 100 \gev\ (Table \ref{tab:Table1}).

\section{Correction to Particle Level}

The inclusive jet distribution is corrected to the particle level. No correction is made for hadronization effects that may modify the energy in the jet cone at the particle level relative to the parton level. This choice is made to facilitate future comparison to jet measurements in heavy-ion collisions, where correction to the parton level is not well-defined at present. 

\begin{figure}[htbp]
\centering
\includegraphics[width=1\textwidth]{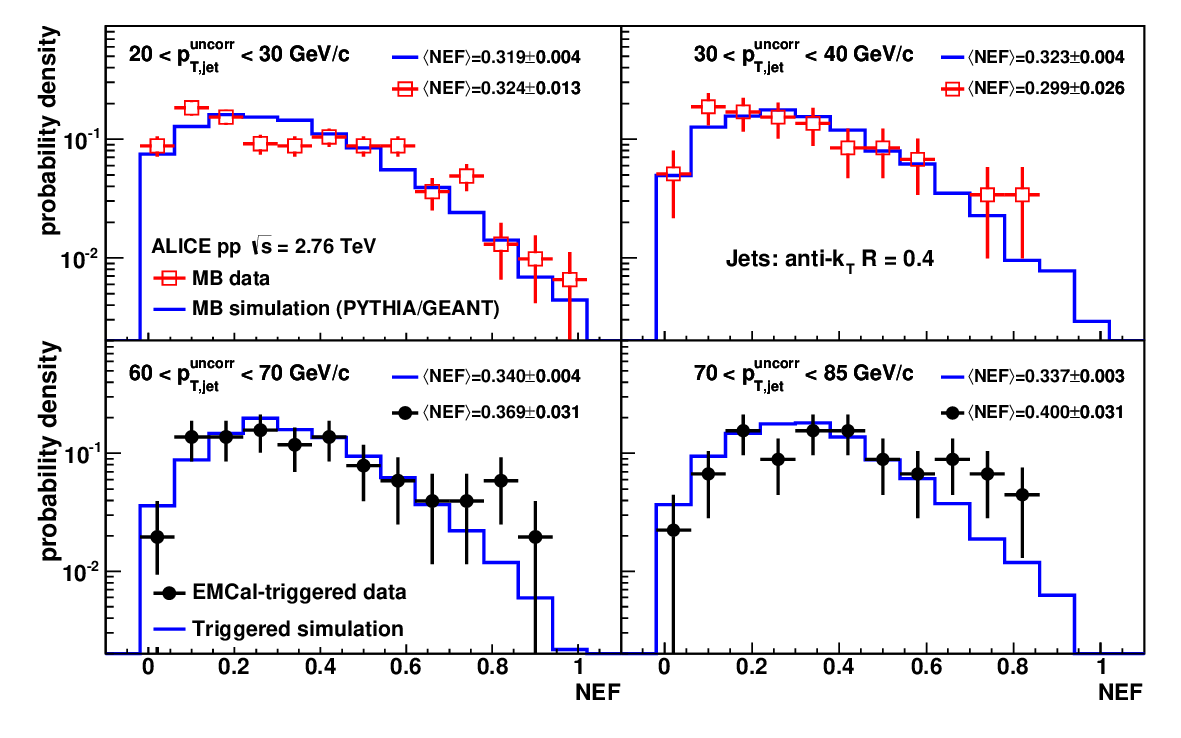}
\caption{Jet neutral energy fraction (NEF) distributions for MB data (open squares), EMCal-triggered data (filled circles) and simulations (histograms), in four different \pTjet\ intervals.
\label{fig::NEF}
}
\end{figure}

Correction is based on detailed, detector-level simulations utilizing PYTHIA6 and GEANT3, which have been validated extensively using ALICE measurements of jets and inclusive particle production. An example of this validation is given in Fig. \ref{fig::NEF}, which shows the distribution of jet neutral energy fraction (NEF) for data and detector-level simulations, for various intervals of \pTjet. Good agreement between data and simulation is observed. Similar levels of agreement are achieved for other key comparisons of data and simulation, including the number of charged track and EMCal cluster constituents per jet, the \zleading\ distribution, the mean \pT\ of clusters and tracks in jets, and the inclusive distributions of identified hadrons over a wide \pT\ range. 
Corrections to the inclusive jet yield are applied bin-by-bin \cite{Cowan}, with correction factor for each bin defined as

\begin{equation}
\CMCwithArgs = \frac
{\int_{\pTlow}^{\pThigh} \mathrm{d} \pT \dfmeasdpT \cdot \frac{\dsigMCpart}{\dsigMCdet}}
{\int_{\pTlow}^{\pThigh} \mathrm{d} \pT \dfmeasdpT},
\label{eq:CMC}
\end{equation}
\noindent
where \dsigMCpart\ and \dsigMCdet\ are the particle-level and detector-level inclusive jet spectra from PYTHIA6; \dfmeasdpT\ is a parametrization of the measured, uncorrected inclusive jet distribution, which provides a weight function to minimize the dependence on the spectral shape of the simulation; and \pTlow\ and \pThigh\ are the bin limits. 

\begin{figure}[htbp]
\centering
\includegraphics[width=0.99\textwidth]{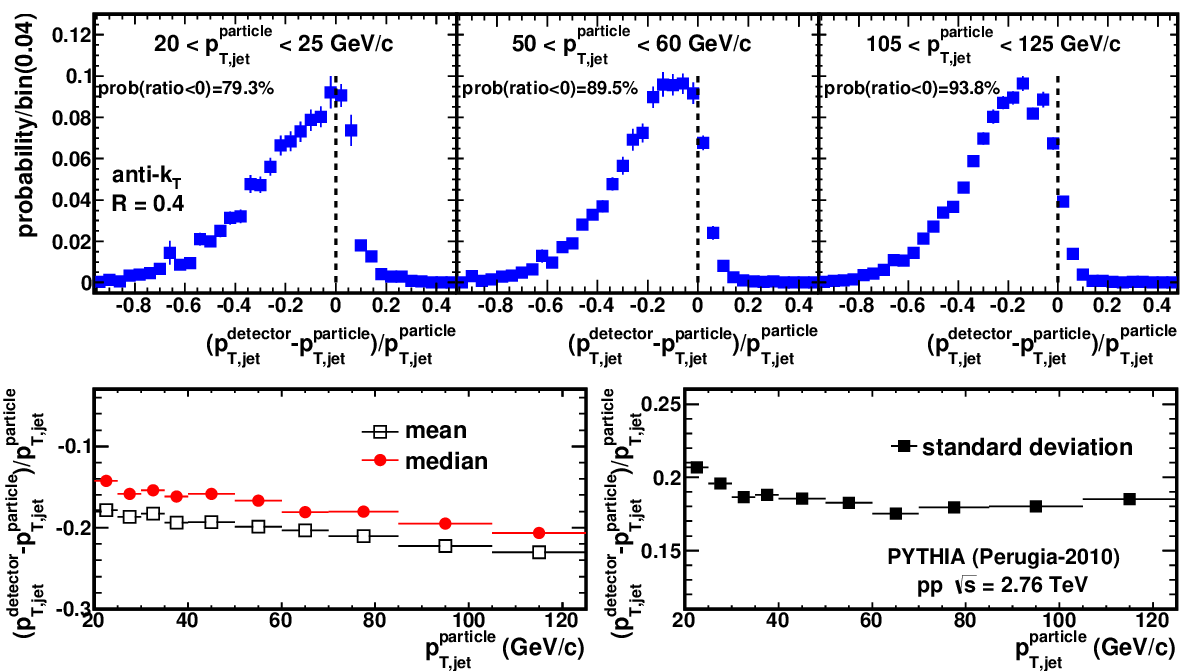}
\caption{Simulated detector response to jets. Upper panels: probability distribution of $\left(\pTdet-\pTpart\right)/\pTpart$, in intervals of \pTpart. Lower left: mean and median of distribution as a function of \pTpart. Lower right: standard deviation of distribution as a function of \pTpart. The statistical errors are smaller than the marker size.
\label{fig:DetectorResponse}}
\end{figure}

Figure \ref{fig:DetectorResponse} illustrates the detector response to jets from simulations, by comparing jet \pT\ at the particle level (\pTpart) and detector level (\pTdet) on a jet-by-jet basis. The upper panels show the probability distribution of their relative difference, for representative intervals in \pTpart. In all cases, \pTdet\ is smaller than \pTpart\ with high probability. This occurs because the largest detector-level effects are due to unobserved particles, i.e. finite charged particle tracking efficiency and undetected neutrons and \kzerol. Correction to jet energy for unmeasured neutron and \kzerol\ energy is estimated to be $3.6-6$\%, depending on jet \pT\ and \rr. Simulation of this component of the particle spectrum was validated by comparison with ALICE measurements of the inclusive spectrum of protons and kaons in \pp\ collisions at \sqrts\ = 2.76 TeV for $\pT<20$ \gev.
Large upward fluctuations in the detector response ($\pTdet>\pTpart$), which are much less probable, are due predominantly to rare track configurations in which daughters of secondary vertices are incorrectly reconstructed with high \pT, with their contribution not eliminated by the cuts described above. Comparison of simulations and data show that these configurations are accurately modeled in the simulations. Their rate in data is small and they make negligible contribution to the measured jet spectrum.

Figure \ref{fig:DetectorResponse}, lower left, shows the mean and median of the relative difference between \pTpart\ and \pTdet, as a function of \pTpart. The median correction to the jet energy is about 15\% at \pT\ = 25 \gev\ and 19\% at
\pT\ = 100 \gev. Figure \ref{fig:DetectorResponse}, lower right, shows the standard deviation of the relative difference as a function of \pTpart, corresponding to an estimate of Jet Energy Resolution (JER) approximately 18\%. However, the distributions in the upper panels are seen to be significantly non-Gaussian, especially at low \pTpart, so that the median shift and standard deviation do not fully characterize the detector response. The full distribution of the detector response is used to determine \dsigMCdet\ in \CMC. For \rr\ = 0.4, \CMC\ rises monotonically from 1.5 at \pT\ = 20 \gev\ to 2.5 at \pT\ = 120 \gev, while for \rr\ = 0.2, \CMC\ rises monotonically from 1.7 at \pT\ = 20 \gev\ to 2.7 at \pT\ = 120 \gev.

Table \ref{tab:Table1} shows all contributions to the systematic uncertainty of JES, determined from the variation in corrected jet yield arising from systematic variation of components of the detector response and analysis algorithms. A given fractional variation in JES corresponds to a fractional variation in jet yield approximately five times larger. The uncertainty due to unmeasured neutron and \kzerol\ energy is estimated by comparing the corrections based on PYTHIA6 and HERWIG. The EMCal energy scale uncertainty is determined by comparing the $\pi^{\rm{0}}$ mass position and $E/p$ of electrons between data and simulation. Systematic sensitivity to the EMCal clustering algorithm is explored with an alternative approach, in which clusters are strictly limited to $3\times3$ adjacent EMCal towers, resulting in 1\% systematic uncertainty. The systematic uncertainty due to the EMCal cluster non-linearity correction is assessed by omitting this correction in the analysis. The systematic uncertainty due to the correction for charged hadron energy deposition in the EMCal is estimated by varying both \fsub\ and the track-cluster matching criteria. Sensitivity to the relative contribution of quark and gluon jets is assessed by tagging each jet from PYTHIA6 according to the highest energy parton within its phase space, and calculating \CMC\ separately for quark and gluon-initiated jets. PYTHIA6 estimates that gluon-initiated jets make up about 70\% of the jet population within the kinematic region of this measurement, and variation of the $q/g$ ratio by 10\% relative to that in PYTHIA6 contributes 1\% uncertainty to the fragmentation model dependence of JES. The total JES systematic uncertainty is less than 3.6\%.


\begin{table}[tp] \caption{Systematic uncertainty of the Jet Energy Scale (JES). Data at 25 \gev\ are from the MB data set, whereas data at 100 \gev\ are from the EMCal-triggered data set. }
\label{tab:Table1}
\centering 
\begin{tabular}{ccccc}
  \hline
  \hline
  &\multicolumn{2}{c}{Jets \rr\ = 0.2} & \multicolumn{2}{c}{Jets \rr\ = 0.4}\\  [-0.1ex]
  \cline{2-3}\cline{4-5}
  \raisebox{1ex}{Source of systematic uncertainty} & 25 \gev\ & 100 \gev\ & 25 \gev\ & 100 \gev \\ [0.5ex]
  \hline
Tracking efficiency & $1.4\%$ & $2.2\%$ & $1.8\%$ & $2.4\%$ \\
Momentum scale of charged tracks & negligible & negligible & negligible & negligible \\
Charged hadron showering in EMCal& $0.7\%$ & $1.4\%$& $0.6\%$ & $1.6\%$ \\
Energy scale of EMCal cluster & $0.6\%$ & $0.6\%$ & $0.8\%$ & $0.8\%$\\
EMCal non-linearity & $0.3\%$ & negligible & $0.6\%$ & negligible \\
EMCal clustering algorithm & $1.0\%$ & $1.0\%$ & $1.0\%$ & $1.0\%$ \\
Underlying event & $0.2\%$ & negligible & $1.0\%$ & $0.3\%$\\
Unmeasured neutron+\kzerol & $0.6\%$ & $0.6\%$ & $0.6\%$ & $0.6\%$\\
Fragmentation model dependence & $1.6\%$ & $1.6\%$ & $1.6\%$ & $1.6\%$\\
\hline
Total JES uncertainty & $2.6\%$ & $3.3\%$ & $3.1\%$ & $3.6\%$\\
\hline\hline
\end{tabular}
\end{table}


Table \ref{tab:Table2} presents the components of the systematic uncertainty of \CMC\ (Eq.\ \ref{eq:CMC}). The uncertainty due to the particle-level spectrum shape is estimated by fitting the particle-level spectrum with a power law function $\propto1/\pT^n$ ($n\approx5$) and varying $n$ by $\pm 0.5$, which covers the variation in $n$ derived from different Monte Carlo models. The uncertainties due to momentum resolution of charged tracks and energy resolution of the EMCal clusters are estimated from comparison of data and simulations. 

Systematic uncertainties at different \pT\ are largely correlated. The components are added in quadrature to generate the cumulative uncertainty, which is labeled ``Spectrum total systematic uncertainty'' in Table \ref{tab:Table2}, and ``Systematic uncertainty'' in Fig. \ref{fig:JetXSec} and \ref{fig:JetXSecRatio}.


\begin{table}[tp] \caption{Systematic uncertainty of corrections to the inclusive jet cross section. Data at 25 \gev\ are from the MB data set, whereas data at 100 \gev\ are from the EMCal-triggered data set. The values refer to percent variation of the cross section.}
\label{tab:Table2}
\centering 
\begin{tabular}{ccccc}
  \hline
  \hline
  &\multicolumn{2}{c}{Jets \rr\ = 0.2} & \multicolumn{2}{c}{Jets \rr\ = 0.4}\\  [-0.1ex]
  \cline{2-3}\cline{4-5}
  \raisebox{1ex}{Sources of systematic uncertainties} & 25 \gev\ & 100 \gev\ & 25 \gev\ & 100 \gev \\ [0.5ex]
  \hline 
JES & $13.1\%$ & $16.5\%$ & $15.5\%$ & $18.0\%$\\
Input PYTHIA6 spectrum shape & $4\%$ & $6\%$ & $4\%$ & $7\%$ \\
Momentum resolution of charged track & $2\%$& $2\%$& $3\%$& $3\%$\\
Energy resolution of EMCal cluster & $1\%$& $1\%$ & $1\%$ & $1\%$ \\ 
EMCal-SSh trigger efficiency & none& $1.7\%$& none& $1.8\%$\\
Cross section normalization & $1.9\%$& $1.9\%$& $1.9\%$& $1.9\%$\\
\hline
Spectrum total systematic uncertainty & $14\%$ & $18\%$ &$ 16\%$ &$ 20\%$ \\ 
\hline\hline
\end{tabular}
\end{table}

\section{Results}

\begin{figure}[htbp]
\centering
\includegraphics[width=0.49\textwidth]{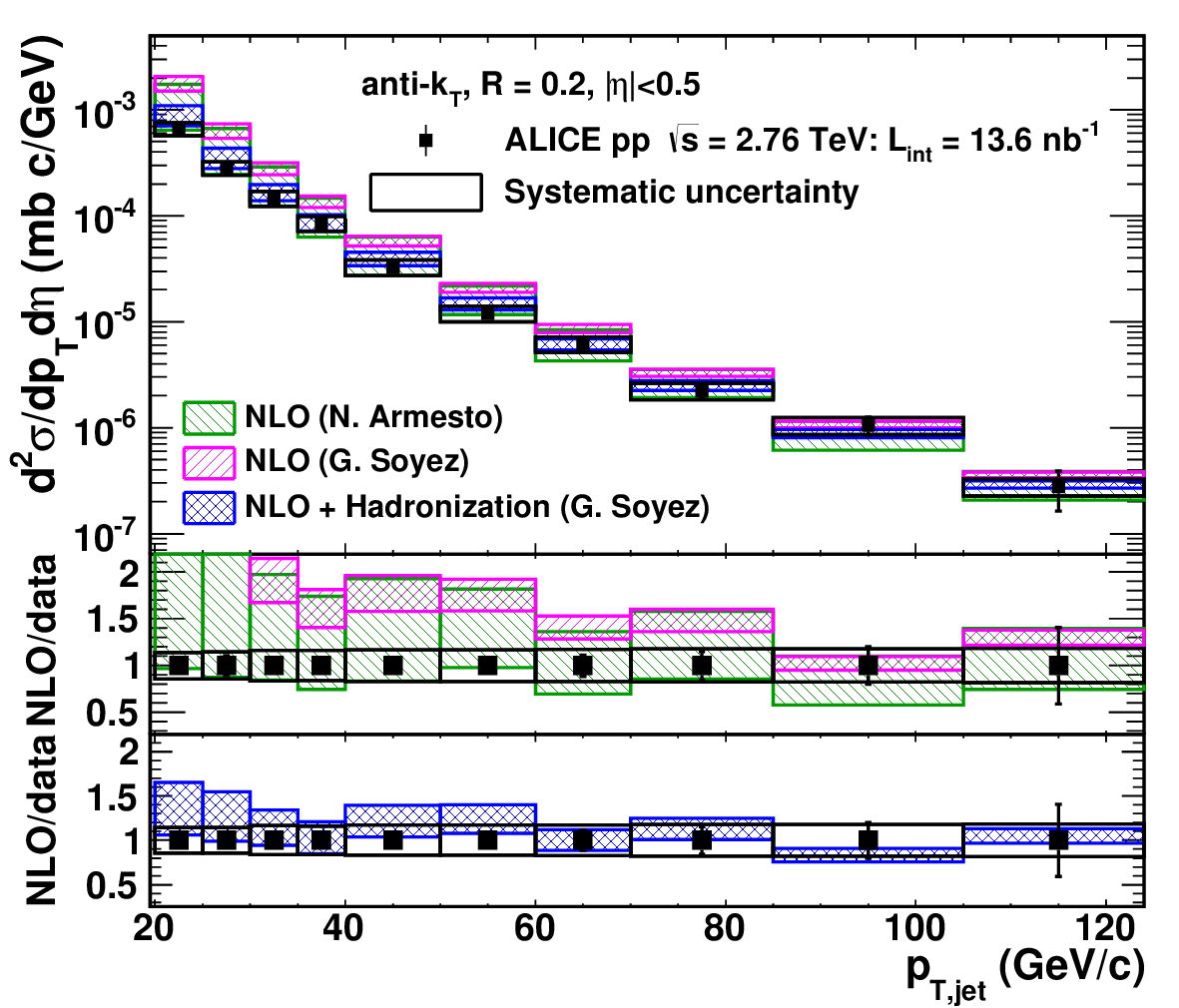}
\includegraphics[width=0.49\textwidth]{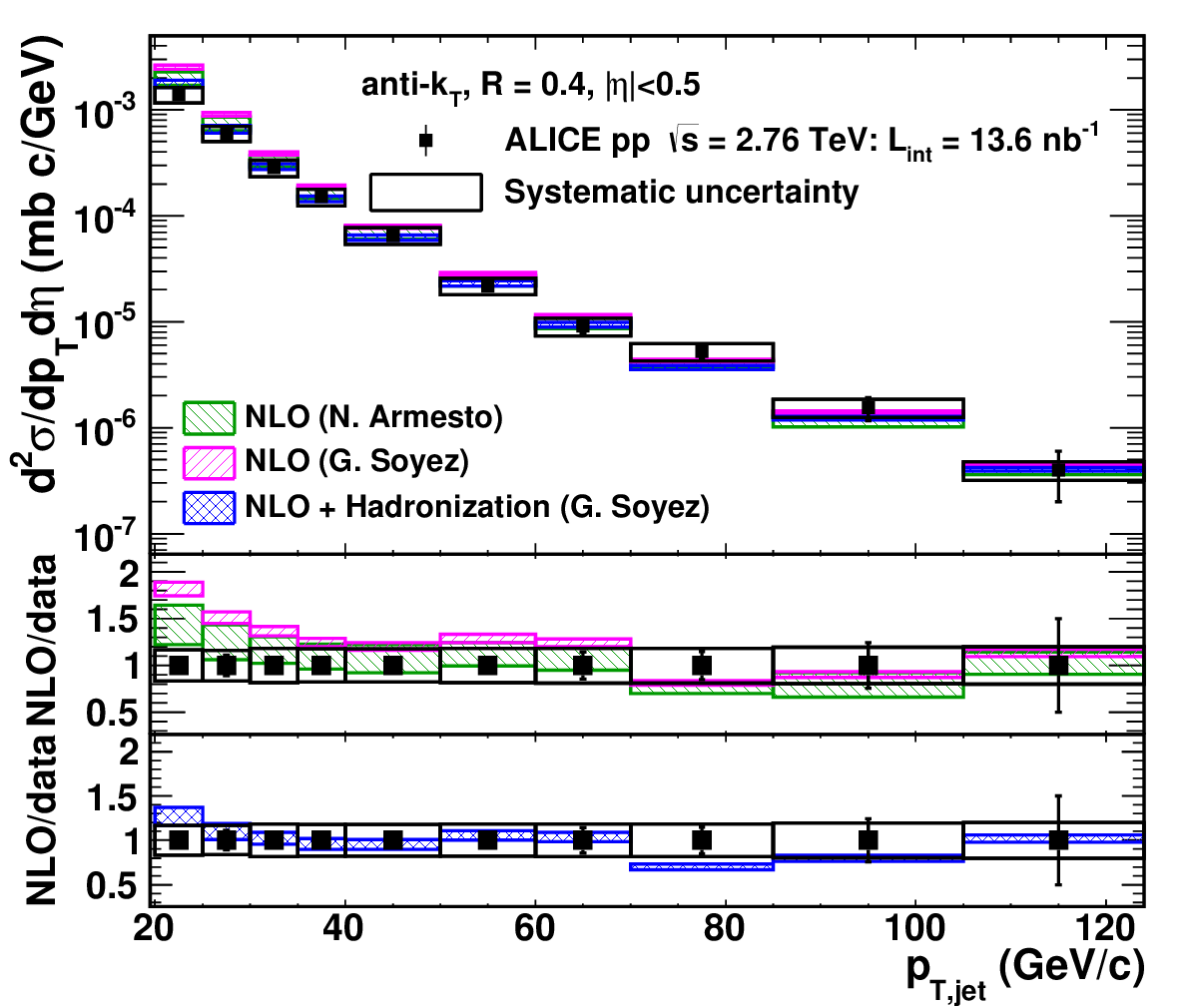}
\caption{Upper panels: inclusive differential jet cross sections for \rr\ = 0.2 (left) and \rr\ = 0.4 (right). Vertical bars show the statistical error, while boxes show the systematic uncertainty (Table \ref{tab:Table2}). The bands show the NLO pQCD calculations discussed in the text \cite{Armesto,Soyez:2011np}. Lower panels: ratio of NLO pQCD calculations to data. Data points are placed at the center of each bin.
\label{fig:JetXSec}
}
\end{figure}

Figure \ref{fig:JetXSec} shows the inclusive differential jet cross-section at particle level for \rr\ = 0.2 (left) and \rr\ = 0.4 (right), together with the results of pQCD calculations at NLO. In order to limit sensitivity to the systematic uncertainty of the SSh trigger efficiency, MB data are used for $\pT<30$ \gev, whereas EMCal-triggered data are used for $\pT>30$ \gev. The Armesto calculation \cite{Armesto} is carried out at the parton level using MSTW08 parton distribution functions (pdf) \cite{Martin:2009iq}. The Soyez calculation utilizes CTEQ6.6 pdfs \cite{Nadolsky:2008zw} and is carried out at both the partonic and hadronization levels \cite{Soyez:2011np}. The bands indicate the theoretical uncertainty estimated by varying the renormalisation and factorisation scales between 0.5 \pT\ to 2.0 \pT. The lower panels of Fig. \ref{fig:JetXSec} show the ratio of the NLO pQCD calculations to data. The calculations for both \rr\ = 0.2 and \rr\ = 0.4 are seen to agree with data within uncertainties, when hadronization effects are included. Both calculations also agree well with inclusive jet cross section measurements at \sqrts\ = 7 TeV \cite{Armesto,Soyez:2011np}.

\begin{figure}[htbp]
\centering
\includegraphics[width=0.6\textwidth]{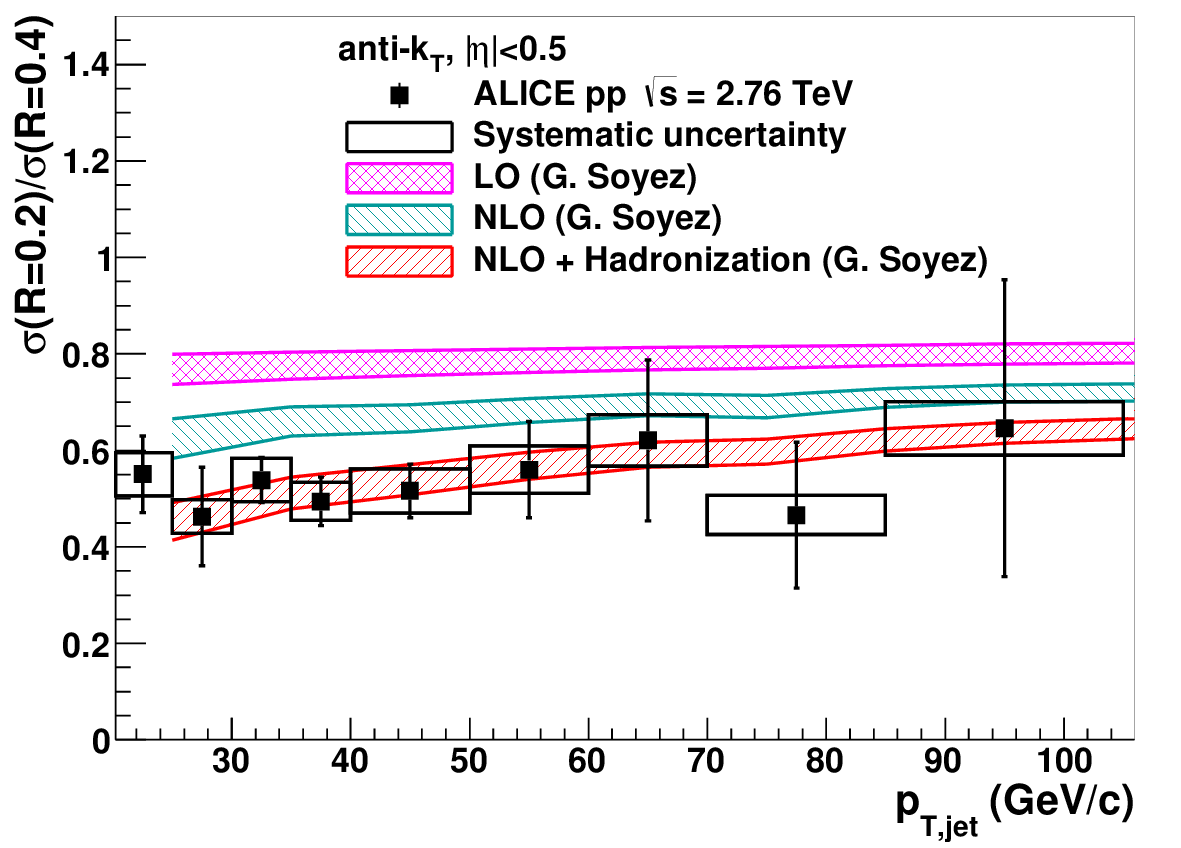}
\caption{Ratio of inclusive differential jet cross sections for \rr\ = 0.2 and \rr\ = 0.4, with pQCD calculations from \cite{Soyez:2011np}. Data points are placed at the center of each bin.
\label{fig:JetXSecRatio}
}
\end{figure}

Figure \ref{fig:JetXSecRatio} shows the ratio of the measured inclusive differential jet cross sections for \rr\ = 0.2 and \rr\ = 0.4. The numerator and denominator utilize disjoint subsets of the data, to ensure that they are statistically independent. The kinematic reach of this measurement is therefore less than that of the individual inclusive spectra. The figure also shows parton-level pQCD calculations at Leading-Order (LO), NLO and NLO with hadronization correction\cite{Soyez:2011np}. This ratio allows a more stringent comparison of data and calculations than the individual inclusive cross sections, since systematic uncertainties that are common or highly correlated, most significantly trigger efficiency, tracking efficiency, and cross section normalization, make smaller relative contribution to the uncertainty of the ratio. In addition, the pQCD calculation considers the ratio directly, rather than each distribution separately, making the calculated ratio effectively one perturbative order higher than the individual cross sections (e.g. the curve labelled ``NLO'' is effectively NNLO) \cite{Soyez:2011np}. 

This ratio, which provides a measurement of the transverse structure of jets, is seen to be less than unity, i.e. at fixed \pT\ the cross section is smaller for \rr\ = 0.2 than for \rr\ = 0.4. The NLO calculation of the ratio agrees within uncertainties with the measurement if hadronization effects are taken into account, indicating that the distribution of radiation within the jet is well-described by the calculation. The transverse structure of jets produced in \pp\ collisions has also been studied using the jet energy profile~\cite{Acosta:2005ix}, whose measurement is described well by a pQCD calculation at NLO with resummation~\cite{Li:2011hy}. Both the cross section ratio presented here and the jet energy profile will be applied in future study of jet quenching in heavy ion collisions.

\section{Summary}
In summary, we have presented the first measurement of the inclusive differential jet cross section at mid-rapidity in \pp\ collisions at \sqrts\ = 2.76 TeV. These data provide an important reference for jet measurements in heavy-ion collisions at the same \sqrtsNN, as well as a test of pQCD calculations at a previously unexamined energy. NLO pQCD calculations with hadronization agree well with both inclusive jet cross section measurements at \rr\ = 0.2 and \rr\ = 0.4, as well as their ratio.

\newenvironment{acknowledgement}{\relax}{\relax}
\begin{acknowledgement}
\section*{Acknowledgements}
We congratulate the LHC for its excellent performance during the special pp run at $\sqrt{s}=2.76$ TeV that generated these data. We thank Gregory Soyez for providing theoretical calculations. 

The ALICE collaboration would like to thank all its engineers and technicians for their invaluable contributions to the construction of the experiment and the CERN accelerator teams for the outstanding performance of the LHC complex.
\\
The ALICE collaboration acknowledges the following funding agencies for their support in building and
running the ALICE detector:
 \\
State Committee of Science, Calouste Gulbenkian Foundation from
Lisbon and Swiss Fonds Kidagan, Armenia;
 \\
Conselho Nacional de Desenvolvimento Cient\'{\i}fico e Tecnol\'{o}gico (CNPq), Financiadora de Estudos e Projetos (FINEP),
Funda\c{c}\~{a}o de Amparo \`{a} Pesquisa do Estado de S\~{a}o Paulo (FAPESP);
 \\
National Natural Science Foundation of China (NSFC), the Chinese Ministry of Education (CMOE)
and the Ministry of Science and Technology of China (MSTC);
 \\
Ministry of Education and Youth of the Czech Republic;
 \\
Danish Natural Science Research Council, the Carlsberg Foundation and the Danish National Research Foundation;
 \\
The European Research Council under the European Community's Seventh Framework Programme;
 \\
Helsinki Institute of Physics and the Academy of Finland;
 \\
French CNRS-IN2P3, the `Region Pays de Loire', `Region Alsace', `Region Auvergne' and CEA, France;
 \\
German BMBF and the Helmholtz Association;
\\
General Secretariat for Research and Technology, Ministry of
Development, Greece;
\\
Hungarian OTKA and National Office for Research and Technology (NKTH);
 \\
Department of Atomic Energy and Department of Science and Technology of the Government of India;
 \\
Istituto Nazionale di Fisica Nucleare (INFN) and Centro Fermi -
Museo Storico della Fisica e Centro Studi e Ricerche "Enrico
Fermi", Italy;
 \\
MEXT Grant-in-Aid for Specially Promoted Research, Ja\-pan;
 \\
Joint Institute for Nuclear Research, Dubna;
 \\
National Research Foundation of Korea (NRF);
 \\
CONACYT, DGAPA, M\'{e}xico, ALFA-EC and the HELEN Program (High-Energy physics Latin-American--European Network);
 \\
Stichting voor Fundamenteel Onderzoek der Materie (FOM) and the Nederlandse Organisatie voor Wetenschappelijk Onderzoek (NWO), Netherlands;
 \\
Research Council of Norway (NFR);
 \\
Polish Ministry of Science and Higher Education;
 \\
National Authority for Scientific Research - NASR (Autoritatea Na\c{t}ional\u{a} pentru Cercetare \c{S}tiin\c{t}ific\u{a} - ANCS);
 \\
Ministry of Education and Science of Russian Federation,
International Science and Technology Center, Russian Academy of
Sciences, Russian Federal Agency of Atomic Energy, Russian Federal
Agency for Science and Innovations and CERN-INTAS;
 \\
Ministry of Education of Slovakia;
 \\
Department of Science and Technology, South Africa;
 \\
CIEMAT, EELA, Ministerio de Educaci\'{o}n y Ciencia of Spain, Xunta de Galicia (Conseller\'{\i}a de Educaci\'{o}n),
CEA\-DEN, Cubaenerg\'{\i}a, Cuba, and IAEA (International Atomic Energy Agency);
 \\
Swedish Research Council (VR) and Knut $\&$ Alice Wallenberg
Foundation (KAW);
 \\
Ukraine Ministry of Education and Science;
 \\
United Kingdom Science and Technology Facilities Council (STFC);
 \\
The United States Department of Energy, the United States National
Science Foundation, the State of Texas, and the State of Ohio.

\end{acknowledgement}





\begin{thebibliography}{10}

\bibitem{Aad:2011fc}
{\bfseries ATLAS Collaboration}, G.~Aad {\em et~al.},
  ``{Measurement of inclusive jet and dijet production in $pp$ collisions at
  $\sqrt{s}=7$ TeV using the ATLAS detector}'',
  \href{http://dx.doi.org/10.1103/PhysRevD.86.014022}{{\em Phys.Rev.}
  {\bfseries D86} (2012) 014022},
\href{http://arxiv.org/abs/1112.6297}{{\ttfamily arXiv:1112.6297 [hep-ex]}}.

\bibitem{CMS:2011ab}
{\bfseries CMS Collaboration}, S.~Chatrchyan {\em et~al.},
  ``{Measurement of the Inclusive Jet Cross Section in $pp$ Collisions at
  $\sqrt{s}=7$ TeV}'',
  \href{http://dx.doi.org/10.1103/PhysRevLett.107.132001}{{\em Phys.Rev.Lett.}
  {\bfseries 107} (2011) 132001},
\href{http://arxiv.org/abs/1106.0208}{{\ttfamily arXiv:1106.0208 [hep-ex]}}.

\bibitem{Abulencia:2007ez}
{\bfseries CDF Collaboration}, A.~Abulencia {\em et~al.},
  ``{Measurement of the Inclusive Jet Cross Section using the {\boldmath
  $k_{\rm T}$} algorithmin {\boldmath $p\overline{p}$} Collisions at{\boldmath
  $\sqrt{s}$} = 1.96 TeV with the CDF II Detector},''
  \href{http://dx.doi.org/10.1103/PhysRevD.75.119901,
  10.1103/PhysRevD.75.092006}{{\em Phys.Rev.} {\bfseries D75} (2007) 092006},
\href{http://arxiv.org/abs/hep-ex/0701051}{{\ttfamily arXiv:hep-ex/0701051
  [hep-ex]}}.

\bibitem{Aaltonen:2008eq}
{\bfseries CDF Collaboration}, T.~Aaltonen {\em et~al.},
  ``{Measurement of the Inclusive Jet Cross Section at the Fermilab Tevatron p
  anti-p Collider Using a Cone-Based Jet Algorithm},''
  \href{http://dx.doi.org/10.1103/PhysRevD.79.119902,
  10.1103/PhysRevD.78.052006}{{\em Phys.Rev.} {\bfseries D78} (2008) 052006},
\href{http://arxiv.org/abs/0807.2204}{{\ttfamily arXiv:0807.2204 [hep-ex]}}.

\bibitem{Abazov:2008ae}
{\bfseries D0 Collaboration}, V.~Abazov {\em et~al.},
  ``{Measurement of the inclusive jet cross-section in $p \bar{p}$ collisions
  at $\sqrt{s}$ =1.96 TeV},''
  \href{http://dx.doi.org/10.1103/PhysRevLett.101.062001}{{\em Phys.Rev.Lett.}
  {\bfseries 101} (2008) 062001},
\href{http://arxiv.org/abs/0802.2400}{{\ttfamily arXiv:0802.2400 [hep-ex]}}.

\bibitem{Majumder:2010qh}
A.~Majumder and M.~Van~Leeuwen, ``{The Theory and Phenomenology of Perturbative
  QCD Based Jet Quenching}'',
  \href{http://dx.doi.org/10.1016/j.ppnp.2010.09.001}{{\em
  Prog.Part.Nucl.Phys.} {\bfseries A66} (2011) 41--92},
\href{http://arxiv.org/abs/1002.2206}{{\ttfamily arXiv:1002.2206 [hep-ph]}}.

\bibitem{Cacciari:2008gp}
M.~Cacciari, G.~P. Salam, and G.~Soyez, ``{The Anti-kt jet clustering
  algorithm}'', \href{http://dx.doi.org/10.1088/1126-6708/2008/04/063}{{\em
  JHEP} {\bfseries 0804} (2008) 063},
\href{http://arxiv.org/abs/0802.1189}{{\ttfamily arXiv:0802.1189 [hep-ph]}}.

\bibitem{Cacciari:2011ma}
M.~Cacciari, G.~P. Salam, and G.~Soyez, ``{FastJet User Manual}'',
  \href{http://dx.doi.org/10.1140/epjc/s10052-012-1896-2}{{\em Eur.Phys.J.}
  {\bfseries C72} (2012) 1896},
\href{http://arxiv.org/abs/1111.6097}{{\ttfamily arXiv:1111.6097 [hep-ph]}}.

\bibitem{Frixione:1995ms}
S.~Frixione, Z.~Kunszt, and A.~Signer, ``{Three jet cross-sections to
  next-to-leading order}'',
  \href{http://dx.doi.org/10.1016/0550-3213(96)00110-1}{{\em Nucl.Phys.}
  {\bfseries B467} (1996) 399--442},
\href{http://arxiv.org/abs/hep-ph/9512328}{{\ttfamily arXiv:hep-ph/9512328
  [hep-ph]}}.

\bibitem{Frixione:1997np}
S.~Frixione, ``{A General approach to jet cross-sections in QCD}'',
  \href{http://dx.doi.org/10.1016/S0550-3213(97)00574-9}{{\em Nucl.Phys.}
  {\bfseries B507} (1997) 295--314},
\href{http://arxiv.org/abs/hep-ph/9706545}{{\ttfamily arXiv:hep-ph/9706545
  [hep-ph]}}.

\bibitem{Armesto}
N.~Armesto, Private communication. Calculations based on
  \cite{Frixione:1995ms,Frixione:1997np}.

\bibitem{Soyez:2011np}
G.~Soyez, ``{A Simple description of jet cross-section ratios}'',
  \href{http://dx.doi.org/10.1016/j.physletb.2011.02.061}{{\em Phys.Lett.}
  {\bfseries B698} (2011) 59--62},
\href{http://arxiv.org/abs/1101.2665}{{\ttfamily arXiv:1101.2665 [hep-ph]}}.

\bibitem{Aamodt:2008zz}
{\bfseries ALICE Collaboration}, K.~Aamodt {\em et~al.}, ``{The
  ALICE experiment at the CERN LHC}'',
\href{http://dx.doi.org/10.1088/1748-0221/3/08/S08002}{{\em JINST} {\bfseries
  3} (2008) S08002}.

\bibitem{Aamodt:2010aa}
{\bfseries ALICE Collaboration}, K.~Aamodt {\em et~al.},
  ``{Alignment of the ALICE Inner Tracking System with cosmic-ray tracks}'',
  \href{http://dx.doi.org/10.1088/1748-0221/5/03/P03003}{{\em JINST} {\bfseries
  5} (2010) P03003},
\href{http://arxiv.org/abs/1001.0502}{{\ttfamily arXiv:1001.0502
  [physics.ins-det]}}.

\bibitem{Cortese:2008zza}
{\bfseries ALICE Collaboration}, P.~Cortese {\em et~al.},
  ``{ALICE electromagnetic calorimeter technical design report}'',
{\em CERN-LHCC-2008-014} (2008) .

\bibitem{Allen:2009aa}
{\bfseries ALICE EMCal Collaboration}, J.~Allen {\em et~al.},
  ``{Performance of prototypes for the ALICE electromagnetic calorimeter}'',
  \href{http://dx.doi.org/10.1016/j.nima.2009.12.061}{{\em Nucl.Instrum.Meth.}
  {\bfseries A615} (2010) 6--13},
\href{http://arxiv.org/abs/0912.2005}{{\ttfamily arXiv:0912.2005
  [physics.ins-det]}}.

\bibitem{2012sja}
{\bfseries ALICE Collaboration}, B.~Abelev {\em et~al.},
  ``{Measurement of inelastic, single- and double-diffraction cross sections in
  proton--proton collisions at the LHC with ALICE}'',
\href{http://arxiv.org/abs/1208.4968}{{\ttfamily arXiv:1208.4968 [hep-ex]}}.

\bibitem{Abelev:2012hxa}
{\bfseries ALICE Collaboration}, B.~Abelev {\em et~al.},
  ``{Centrality Dependence of Charged Particle Production at Large Transverse
  Momentum in Pb--Pb Collisions at $\sqrt{s_{\rm{NN}}} = 2.76$ TeV},''
  \href{http://dx.doi.org/10.1016/j.physletb.2013.01.051}{{\em Phys.Lett.}
  {\bfseries B720} (2013) 52--62},
\href{http://arxiv.org/abs/1208.2711}{{\ttfamily arXiv:1208.2711 [hep-ex]}}.

\bibitem{Sjostrand:2006za}
T.~Sjostrand, S.~Mrenna, and P.~Z. Skands, ``{PYTHIA 6.4 Physics and Manual}'',
  \href{http://dx.doi.org/10.1088/1126-6708/2006/05/026}{{\em JHEP} {\bfseries
  0605} (2006) 026},
\href{http://arxiv.org/abs/hep-ph/0603175}{{\ttfamily arXiv:hep-ph/0603175
  [hep-ph]}}.

\bibitem{HERWIG}
{G.Corcella, I.G.Knowles, G.Marchesini, S.Moretti, K.Odagiri, P.Richardson,
  M.H.Seymour and B.R.Webber}, ``{HERWIG 6: an event generator for Hadron
  Emission Reactions With Interfering Gluons (including supersymmetric
  processes)}'', \href{http://dx.doi.org/10.1088/1126-6708/2001/01/010}{{\em
  JHEP} {\bfseries 0101} (2001) 010},
  \href{http://arxiv.org/abs/0011363v3}{{\ttfamily arXiv:0011363v3 [hep-ph]}}.

\bibitem{GEANT3}
{R. Brun, F. Bruyant, M. Maire, A.C. McPherson, and P. Zanarini}, ``{GEANT3
  User Guide}'', {\em CERN Data Handling Division DD/EE/84-1} (1985) .

\bibitem{Alme:2010ke}
J.~Alme, Y.~Andres, H.~Appelshauser, S.~Bablok, N.~Bialas, {\em et~al.}, ``{The
  ALICE TPC, a large 3-dimensional tracking device with fast readout for
  ultra-high multiplicity events}'',
  \href{http://dx.doi.org/10.1016/j.nima.2010.04.042}{{\em Nucl.Instrum.Meth.}
  {\bfseries A622} (2010) 316--367},
\href{http://arxiv.org/abs/1001.1950}{{\ttfamily arXiv:1001.1950
  [physics.ins-det]}}.

\bibitem{Fruhwirth:1987fm}
R.~Fruhwirth, ``{Application of Kalman filtering to track and vertex
  fitting}'',
\href{http://dx.doi.org/10.1016/0168-9002(87)90887-4}{{\em Nucl.Instrum.Meth.}
  {\bfseries A262} (1987) 444--450}.

\bibitem{CMS:2010eua}
{\bfseries CMS Collaboration},
  ``{Commissioning of the Particle-Flow reconstruction in Minimum-Bias and Jet
  Events from pp Collisions at 7 TeV},''
{\em CMS-PAS-PFT-10-002} (2010) .

\bibitem{CDFUE}
{\bfseries CDF Collaboration}, A.~Cruz {\em et~al.}, ``{Using
  MAX/MIN Transverse Regions to Study the Underlying Event in Run 2 at the
  Tevatron}'',
{\em CDF/ANAL/CDF/CDFR/7703} (2005) .

\bibitem{Cowan}
G.~Cowan, ``{A Survey of Unfolding Methods for Particle Physics}'', {\em {Proc.
  Advanced Statistical Techniques in Particle Physics, Durham}} (2002) .

\bibitem{Martin:2009iq}
A.~Martin, W.~Stirling, R.~Thorne, and G.~Watt, ``{Parton distributions for the
  LHC}'', \href{http://dx.doi.org/10.1140/epjc/s10052-009-1072-5}{{\em
  Eur.Phys.J.} {\bfseries C63} (2009) 189--285},
\href{http://arxiv.org/abs/0901.0002}{{\ttfamily arXiv:0901.0002 [hep-ph]}}.

\bibitem{Nadolsky:2008zw}
P.~M. Nadolsky, H.-L. Lai, Q.-H. Cao, J.~Huston, J.~Pumplin, {\em et~al.},
  ``{Implications of CTEQ global analysis for collider observables}'',
  \href{http://dx.doi.org/10.1103/PhysRevD.78.013004}{{\em Phys.Rev.}
  {\bfseries D78} (2008) 013004},
\href{http://arxiv.org/abs/0802.0007}{{\ttfamily arXiv:0802.0007 [hep-ph]}}.

\bibitem{Acosta:2005ix}
{\bfseries CDF Collaboration}, D.~Acosta {\em et~al.}, ``{Study
  of jet shapes in inclusive jet production in $p\bar{p}$ collisions at
  $\sqrt{s}=1.96$ TeV},''
  \href{http://dx.doi.org/10.1103/PhysRevD.71.112002}{{\em Phys.Rev.}
  {\bfseries D71} (2005) 112002},
\href{http://arxiv.org/abs/hep-ex/0505013}{{\ttfamily arXiv:hep-ex/0505013
  [hep-ex]}}.

\bibitem{Li:2011hy}
H.-n. Li, Z.~Li, and C.-P. Yuan, ``{QCD resummation for jet substructures},''
  \href{http://dx.doi.org/10.1103/PhysRevLett.107.152001}{{\em Phys.Rev.Lett.}
  {\bfseries 107} (2011) 152001},
\href{http://arxiv.org/abs/1107.4535}{{\ttfamily arXiv:1107.4535 [hep-ph]}}.

\end{thebibliography}
\providecommand{\href}[2]{#2}\begingroup\raggedright\endgroup

\newpage
\appendix
\section{The ALICE Collaboration}
\label{app:collab}

\begingroup
\small
\begin{flushleft}
B.~Abelev\Irefn{org1234}\And
J.~Adam\Irefn{org1274}\And
D.~Adamov\'{a}\Irefn{org1283}\And
A.M.~Adare\Irefn{org1260}\And
M.M.~Aggarwal\Irefn{org1157}\And
G.~Aglieri~Rinella\Irefn{org1192}\And
M.~Agnello\Irefn{org1313}\And
A.G.~Agocs\Irefn{org1143}\And
A.~Agostinelli\Irefn{org1132}\And
Z.~Ahammed\Irefn{org1225}
% ...
\renewcommand\labelenumi{\textsuperscript{\theenumi}~}
\section*{Affiliation notes}
\renewcommand\theenumi{\roman{enumi}}
\begin{Authlist}
\item \Adef{0}Deceased
\item \Adef{M.V.Lomonosov Moscow State University, D.V.Skobeltsyn Institute of Nuclear Physics, Moscow, Russia}Also at: M.V.Lomonosov Moscow State University, D.V.Skobeltsyn Institute of Nuclear Physics, Moscow, Russia
\item \Adef{University of Belgrade, Faculty of Physics and Vinvca Institute of Nuclear Sciences, Belgrade, Serbia}Also at: University of Belgrade, Faculty of Physics and Vinvca Institute of Nuclear Sciences, Belgrade, Serbia
\end{Authlist}
\section*{Collaboration Institutes}
\renewcommand\theenumi{\arabic{enumi}~}
\begin{Authlist}
\item \Idef{org1332}A. I. Alikhanyan National Science Laboratory (Yerevan Physics Institute) Foundation, Yerevan, Armenia
\item \Idef{org1279}Benem\'{e}rita Universidad Aut\'{o}noma de Puebla, Puebla, Mexico
\item \Idef{org1220}Bogolyubov Institute for Theoretical Physics, Kiev, Ukraine
\item \Idef{org20959}Bose Institute, Department of Physics and Centre for Astroparticle Physics and Space Science (CAPSS), Kolkata, India
\item \Idef{org1262}Budker Institute for Nuclear Physics, Novosibirsk, Russia
\item \Idef{org1292}California Polytechnic State University, San Luis Obispo, California, United States
\item \Idef{org1329}Central China Normal University, Wuhan, China
\item \Idef{org14939}Centre de Calcul de l'IN2P3, Villeurbanne, France
\item \Idef{org1197}Centro de Aplicaciones Tecnol\'{o}gicas y Desarrollo Nuclear (CEADEN), Havana, Cuba
\item \Idef{org1242}Centro de Investigaciones Energ\'{e}ticas Medioambientales y Tecnol\'{o}gicas (CIEMAT), Madrid, Spain
\item \Idef{org1244}Centro de Investigaci\'{o}n y de Estudios Avanzados (CINVESTAV), Mexico City and M\'{e}rida, Mexico
\item \Idef{org1335}Centro Fermi -- Centro Studi e Ricerche e Museo Storico della Fisica ``Enrico Fermi'', Rome, Italy
\item \Idef{org17347}Chicago State University, Chicago, United States
\item \Idef{org1288}Commissariat \`{a} l'Energie Atomique, IRFU, Saclay, France
\item \Idef{org15782}COMSATS Institute of Information Technology (CIIT), Islamabad, Pakistan
\item \Idef{org1294}Departamento de F\'{\i}sica de Part\'{\i}culas and IGFAE, Universidad de Santiago de Compostela, Santiago de Compostela, Spain
\item \Idef{org1106}Department of Physics Aligarh Muslim University, Aligarh, India
\item \Idef{org1121}Department of Physics and Technology, University of Bergen, Bergen, Norway
\item \Idef{org1162}Department of Physics, Ohio State University, Columbus, Ohio, United States
\item \Idef{org1300}Department of Physics, Sejong University, Seoul, South Korea
\item \Idef{org1268}Department of Physics, University of Oslo, Oslo, Norway
\item \Idef{org1312}Dipartimento di Fisica dell'Universit\`{a} and Sezione INFN, Turin, Italy
\item \Idef{org1132}Dipartimento di Fisica dell'Universit\`{a} and Sezione INFN, Bologna, Italy
\item \Idef{org1145}Dipartimento di Fisica dell'Universit\`{a} and Sezione INFN, Cagliari, Italy
\item \Idef{org1315}Dipartimento di Fisica dell'Universit\`{a} and Sezione INFN, Trieste, Italy
\item \Idef{org1285}Dipartimento di Fisica dell'Universit\`{a} `La Sapienza' and Sezione INFN, Rome, Italy
\item \Idef{org1154}Dipartimento di Fisica e Astronomia dell'Universit\`{a} and Sezione INFN, Catania, Italy
\item \Idef{org1270}Dipartimento di Fisica e Astronomia dell'Universit\`{a} and Sezione INFN, Padova, Italy
\item \Idef{org1290}Dipartimento di Fisica `E.R.~Caianiello' dell'Universit\`{a} and Gruppo Collegato INFN, Salerno, Italy
\item \Idef{org1103}Dipartimento di Scienze e Innovazione Tecnologica dell'Universit\`{a} del Piemonte Orientale and Gruppo Collegato INFN, Alessandria, Italy
\item \Idef{org1114}Dipartimento Interateneo di Fisica `M.~Merlin' and Sezione INFN, Bari, Italy
\item \Idef{org1237}Division of Experimental High Energy Physics, University of Lund, Lund, Sweden
\item \Idef{org1192}European Organization for Nuclear Research (CERN), Geneva, Switzerland
\item \Idef{org1227}Fachhochschule K\"{o}ln, K\"{o}ln, Germany
\item \Idef{org1122}Faculty of Engineering, Bergen University College, Bergen, Norway
\item \Idef{org1136}Faculty of Mathematics, Physics and Informatics, Comenius University, Bratislava, Slovakia
\item \Idef{org1274}Faculty of Nuclear Sciences and Physical Engineering, Czech Technical University in Prague, Prague, Czech Republic
\item \Idef{org1229}Faculty of Science, P.J.~\v{S}af\'{a}rik University, Ko\v{s}ice, Slovakia
\item \Idef{org1184}Frankfurt Institute for Advanced Studies, Johann Wolfgang Goethe-Universit\"{a}t Frankfurt, Frankfurt, Germany
\item \Idef{org1215}Gangneung-Wonju National University, Gangneung, South Korea
\item \Idef{org20958}Gauhati University, Department of Physics, Guwahati, India
\item \Idef{org1212}Helsinki Institute of Physics (HIP) and University of Jyv\"{a}skyl\"{a}, Jyv\"{a}skyl\"{a}, Finland
\item \Idef{org1203}Hiroshima University, Hiroshima, Japan
\item \Idef{org1254}Indian Institute of Technology Bombay (IIT), Mumbai, India
\item \Idef{org36378}Indian Institute of Technology Indore (IIT), Indore, India
\item \Idef{org1266}Institut de Physique Nucl\'{e}aire d'Orsay (IPNO), Universit\'{e} Paris-Sud, CNRS-IN2P3, Orsay, France
\item \Idef{org1277}Institute for High Energy Physics, Protvino, Russia
\item \Idef{org1249}Institute for Nuclear Research, Academy of Sciences, Moscow, Russia
\item \Idef{org1320}Nikhef, National Institute for Subatomic Physics and Institute for Subatomic Physics of Utrecht University, Utrecht, Netherlands
\item \Idef{org1250}Institute for Theoretical and Experimental Physics, Moscow, Russia
\item \Idef{org1230}Institute of Experimental Physics, Slovak Academy of Sciences, Ko\v{s}ice, Slovakia
\item \Idef{org1127}Institute of Physics, Bhubaneswar, India
\item \Idef{org1275}Institute of Physics, Academy of Sciences of the Czech Republic, Prague, Czech Republic
\item \Idef{org1139}Institute of Space Sciences (ISS), Bucharest, Romania
\item \Idef{org27399}Institut f\"{u}r Informatik, Johann Wolfgang Goethe-Universit\"{a}t Frankfurt, Frankfurt, Germany
\item \Idef{org1185}Institut f\"{u}r Kernphysik, Johann Wolfgang Goethe-Universit\"{a}t Frankfurt, Frankfurt, Germany
\item \Idef{org1177}Institut f\"{u}r Kernphysik, Technische Universit\"{a}t Darmstadt, Darmstadt, Germany
\item \Idef{org1256}Institut f\"{u}r Kernphysik, Westf\"{a}lische Wilhelms-Universit\"{a}t M\"{u}nster, M\"{u}nster, Germany
\item \Idef{org1246}Instituto de Ciencias Nucleares, Universidad Nacional Aut\'{o}noma de M\'{e}xico, Mexico City, Mexico
\item \Idef{org1247}Instituto de F\'{\i}sica, Universidad Nacional Aut\'{o}noma de M\'{e}xico, Mexico City, Mexico
\item \Idef{org23333}Institut of Theoretical Physics, University of Wroclaw, Wroclaw, Poland
\item \Idef{org1308}Institut Pluridisciplinaire Hubert Curien (IPHC), Universit\'{e} de Strasbourg, CNRS-IN2P3, Strasbourg, France
\item \Idef{org1182}Joint Institute for Nuclear Research (JINR), Dubna, Russia
\item \Idef{org1143}KFKI Research Institute for Particle and Nuclear Physics, Hungarian Academy of Sciences, Budapest, Hungary
\item \Idef{org1199}Kirchhoff-Institut f\"{u}r Physik, Ruprecht-Karls-Universit\"{a}t Heidelberg, Heidelberg, Germany
\item \Idef{org20954}Korea Institute of Science and Technology Information, Daejeon, South Korea
\item \Idef{org1160}Laboratoire de Physique Corpusculaire (LPC), Clermont Universit\'{e}, Universit\'{e} Blaise Pascal, CNRS--IN2P3, Clermont-Ferrand, France
\item \Idef{org1194}Laboratoire de Physique Subatomique et de Cosmologie (LPSC), Universit\'{e} Joseph Fourier, CNRS-IN2P3, Institut Polytechnique de Grenoble, Grenoble, France
\item \Idef{org1187}Laboratori Nazionali di Frascati, INFN, Frascati, Italy
\item \Idef{org1232}Laboratori Nazionali di Legnaro, INFN, Legnaro, Italy
\item \Idef{org1125}Lawrence Berkeley National Laboratory, Berkeley, California, United States
\item \Idef{org1234}Lawrence Livermore National Laboratory, Livermore, California, United States
\item \Idef{org1251}Moscow Engineering Physics Institute, Moscow, Russia
\item \Idef{org1322}National Centre for Nuclear Studies, Warsaw, Poland
\item \Idef{org1140}National Institute for Physics and Nuclear Engineering, Bucharest, Romania
\item \Idef{org1165}Niels Bohr Institute, University of Copenhagen, Copenhagen, Denmark
\item \Idef{org1109}Nikhef, National Institute for Subatomic Physics, Amsterdam, Netherlands
\item \Idef{org1283}Nuclear Physics Institute, Academy of Sciences of the Czech Republic, \v{R}e\v{z} u Prahy, Czech Republic
\item \Idef{org1264}Oak Ridge National Laboratory, Oak Ridge, Tennessee, United States
\item \Idef{org1189}Petersburg Nuclear Physics Institute, Gatchina, Russia
\item \Idef{org1170}Physics Department, Creighton University, Omaha, Nebraska, United States
\item \Idef{org1157}Physics Department, Panjab University, Chandigarh, India
\item \Idef{org1112}Physics Department, University of Athens, Athens, Greece
\item \Idef{org1152}Physics Department, University of Cape Town and  iThemba LABS, National Research Foundation, Somerset West, South Africa
\item \Idef{org1209}Physics Department, University of Jammu, Jammu, India
\item \Idef{org1207}Physics Department, University of Rajasthan, Jaipur, India
\item \Idef{org1200}Physikalisches Institut, Ruprecht-Karls-Universit\"{a}t Heidelberg, Heidelberg, Germany
\item \Idef{org1325}Purdue University, West Lafayette, Indiana, United States
\item \Idef{org1281}Pusan National University, Pusan, South Korea
\item \Idef{org1176}Research Division and ExtreMe Matter Institute EMMI, GSI Helmholtzzentrum f\"ur Schwerionenforschung, Darmstadt, Germany
\item \Idef{org1334}Rudjer Bo\v{s}kovi\'{c} Institute, Zagreb, Croatia
\item \Idef{org1298}Russian Federal Nuclear Center (VNIIEF), Sarov, Russia
\item \Idef{org1252}Russian Research Centre Kurchatov Institute, Moscow, Russia
\item \Idef{org1224}Saha Institute of Nuclear Physics, Kolkata, India
\item \Idef{org1130}School of Physics and Astronomy, University of Birmingham, Birmingham, United Kingdom
\item \Idef{org1338}Secci\'{o}n F\'{\i}sica, Departamento de Ciencias, Pontificia Universidad Cat\'{o}lica del Per\'{u}, Lima, Peru
\item \Idef{org1133}Sezione INFN, Bologna, Italy
\item \Idef{org1155}Sezione INFN, Catania, Italy
\item \Idef{org1271}Sezione INFN, Padova, Italy
\item \Idef{org1146}Sezione INFN, Cagliari, Italy
\item \Idef{org1313}Sezione INFN, Turin, Italy
\item \Idef{org1316}Sezione INFN, Trieste, Italy
\item \Idef{org1286}Sezione INFN, Rome, Italy
\item \Idef{org1115}Sezione INFN, Bari, Italy
\item \Idef{org36377}Nuclear Physics Group, STFC Daresbury Laboratory, Daresbury, United Kingdom
\item \Idef{org1258}SUBATECH, Ecole des Mines de Nantes, Universit\'{e} de Nantes, CNRS-IN2P3, Nantes, France
\item \Idef{org1304}Technical University of Split FESB, Split, Croatia
\item \Idef{org1168}The Henryk Niewodniczanski Institute of Nuclear Physics, Polish Academy of Sciences, Cracow, Poland
\item \Idef{org17361}The University of Texas at Austin, Physics Department, Austin, TX, United States
\item \Idef{org1173}Universidad Aut\'{o}noma de Sinaloa, Culiac\'{a}n, Mexico
\item \Idef{org1296}Universidade de S\~{a}o Paulo (USP), S\~{a}o Paulo, Brazil
\item \Idef{org1149}Universidade Estadual de Campinas (UNICAMP), Campinas, Brazil
\item \Idef{org1239}Universit\'{e} de Lyon, Universit\'{e} Lyon 1, CNRS/IN2P3, IPN-Lyon, Villeurbanne, France
\item \Idef{org1205}University of Houston, Houston, Texas, United States
\item \Idef{org20371}University of Technology and Austrian Academy of Sciences, Vienna, Austria
\item \Idef{org1222}University of Tennessee, Knoxville, Tennessee, United States
\item \Idef{org1310}University of Tokyo, Tokyo, Japan
\item \Idef{org1318}University of Tsukuba, Tsukuba, Japan
\item \Idef{org21360}Eberhard Karls Universit\"{a}t T\"{u}bingen, T\"{u}bingen, Germany
\item \Idef{org1225}Variable Energy Cyclotron Centre, Kolkata, India
\item \Idef{org1306}V.~Fock Institute for Physics, St. Petersburg State University, St. Petersburg, Russia
\item \Idef{org1323}Warsaw University of Technology, Warsaw, Poland
\item \Idef{org1179}Wayne State University, Detroit, Michigan, United States
\item \Idef{org1260}Yale University, New Haven, Connecticut, United States
\item \Idef{org15649}Yildiz Technical University, Istanbul, Turkey
\item \Idef{org1301}Yonsei University, Seoul, South Korea
\item \Idef{org1327}Zentrum f\"{u}r Technologietransfer und Telekommunikation (ZTT), Fachhochschule Worms, Worms, Germany
\end{Authlist}
\endgroup

\end{document}